# Pressure-induced structural transition of ZnO nanocrystals studied with molecular dynamics

Xinwei Dong[1], Feng Liu[1], Yiqun Xie[1], WangZhou Shi[1], Xiang Ye[1,*,†] and J.Z. Jiang[2,3,*]

[1] *Department of Physics, Shanghai Normal University, Shanghai 200234, P.R. China*
[2] *International Center for New-Structured Materials (ICNSM), Zhejiang University and Laboratory of New-Structured Materials, Department of Materials Science & Engineering, Zhejiang University, Hangzhou 310027, P.R. China.*
[3] *School of Materials Science and Engineering, Hefei University of Technology, Hefei 230009, P.R. China*

We have studied the pressure-induced structural transition of ZnO nanocrystals using constant pressure molecular dynamics simulations for finite system. We have observed the transition from the fourfold coordination wurtzite to the sixfold coordination rocksalt structure, and the process of transition is strongly dependent on the morphology of the nanocrystals. It is found that the perfect faceted ZnO nanocrystals undergo wurtzite to rocksalt transition with a perfect fivefold h-MgO structure as the intermediate status. But for the faceted ones without perfect surface structure, as the number of the atoms removed from the ($001$) and ($00\bar{1}$) surface edge increases, the local morphology will become more similar to spherical. The nanocrystal will receive equal stress from every direction and it will be more difficult to compress the structure along only c axis as the perfect faceted ZnO nanocrystal. In this situation, only partial structure experiences intermediate fivefold coordination structure or even no intermediate fivefold coordination structure exists dependent on the surface disorder level.

---

* Email address: yexiang@shnu.edu.cn (Xiang Ye), jiangjz@zju.edu.cn (J.Z. Jiang)
† Tel: (86)21-64322726, Fax: (86)21-64328894

# 1. Introduction

Zinc oxide (ZnO) has long been of research interest these years for its prospects in a range of technological applications in the field of electronics, optics, optoelectronics, catalysis, chemical sensors, and conductive solar cell window layers.[1-3] ZnO exists in a hexagonal wurtzite structure (*B*4) at normal conditions of temperature and pressure. While ZnO occurs as a mineral, the high-pressure phase may be geologically important as a component of the lower mantle.[4] In order to lower their energies through more efficient volume packing, the phase of ZnO bulk transforms to the cubic rocksalt structure (*B*1) at a pressure in the vicinity of 9 GPa,[5, 6] following an increase of coordination number from 4 to 6 and a large volume decrease of about 17%.[7] Several experimental studies have been done to explore the high-pressure stability of ZnO polymorphs,[6, 8-11] as well as the transition thereof associated, leading to various transition models.[12-15]

Nanostructured materials consisting of small crystallites of diameter 1-100 nm, often have novel physical and chemical properties, differing from those of the corresponding bulk materials. Study of nanostructured materials would help us to understand the mechanism of the transition which is hard to learn from bulk materials. Furthermore, ZnO nanomaterial is an important material,[16, 17] that can serve as building blocks for many nanosysterms. Understanding the mechanism of ZnO nanostructured phase transition more clearly is valuable for their potential application. Owe to their much high surface to volume ratio, nanostructures have received different elastic and thermodynamic properties from their bulk counterparts. It has been found that the transition pressures observed for nanocrystals exceed the bulk transition pressures and this has been attributed to the surface effects.[18-29] Recently, it has been reported that transition pressure of 12 nm grain size ZnO nanocrystal is 15.1 GPa, *i.e.* 50% larger than polycrystal bulk material, which is 9.5 GPa.[25] An in-situ transmission electron microscopy study for phase transition from cubic rocksalt to hexagonal wurtzite structure revealed a Peierls distortion path for dense ZnO nanoparticles.[30]

Along with the experimental efforts, molecular dynamics (MD) simulations represent a promising tool for the investigation of the transition mechanisms, whereas this approach offers femtosecond resolution at the atomistic level of detail which is difficult to trace in experiments. MD simulations of the B4 to B1 structural transition for ZnO have been performed,[31-35] allowing mechanisms to be identified directly from the atomic trajectories. In particular, a hexagonal intermediate fivefold coordination structure is occasionally visited for ZnO bulks, as a result of axial compression, which however does not represent a necessary step for the transition.[31] And ZnO existing in B4 and B1 structures can also transform to fivefold coordination structure when it is in the form of thin nanoplates.[36, 37] For ZnO nanowires, the B4 to fivefold coordination structural transition is found to be energetically favorable above a critical tensile stress of 10 GPa in $[01\bar{1}0]$ nanowires.[38] The five coordination structure is also found as the intermediate structure on B4 to B1 structural transition of the ZnO nanorod is also reported.[32] Surface effects are observed to not only influence dedicated nucleation sites but also impose control mechanisms to phase growth and to specific shape transition.[32] Although theoretical and experimental efforts have been made to probe the structural transition of above ZnO nanostructures, at least to our knowledge, a systematic study of structural transition mechanism of ZnO nanocrystals is still lacking. Dose any intermediate structure exist during B4 to B1 structural transition for the ZnO nanocrystals? If yes, what issue will affect the transition path? There are still something unclear to the microscope transition mechanism of ZnO nanocrystals under high pressure.

In this paper we address the above-mentioned questions using the constant-pressure MD simulations for finite systems. The faceted ZnO nanocrystals, which can be seen as part of ZnO nanowires, and the spherical ZnO nanocrystal are chosen as the study objects. In the following, we first present the computational detail in section II, and then results and discussions in section III, finally, summary remarks are given in section IV.

## 2. Materials and methods

In order to simulate the pressure-driven structural transition of ZnO nanocrystals, we use the constant-pressure MD method developed for finite systems.[39] It is parameter-free in this new constant-pressure MD method, and can be used for any system with arbitrary shape in principle. This method is successfully used for the carbon nanotube systems[40-42] and other nanocrystals.[43] In the present study, all the simulations are carried out at 300 K, where the system temperature is controlled by using a Nosé-Hoover thermostat.[44, 45] The equations of motion are solved by using standard MD based on the Verlet-type algorithm. The time step is set as 2.0 fs. The initial configuration of the faceted $Zn_{560}O_{560}$ nanocrystal with well-defined surface structure is obtained by cleaving the bulk lattice along equivalent (100) B4 planes and at (001) and ($00\bar{1}$) planes perpendicular to the [001] direction of the *c* axis. Then the $Zn_{555}O_{555}$, $Zn_{551}O_{551}$ and $Zn_{547}O_{547}$ nanocrystals are obtained by removing several atoms along the edges of the (001) and ($00\bar{1}$) surfaces based on the structure of $Zn_{560}O_{560}$ as shown in Fig. 1. The spherical $Zn_{502}O_{502}$ nanocrystal structure is cut out of bulk ZnO in B4 structure. The size of above five ZnO nanocrystals studied here are all much closed to each other, so we can only consider the surface effect in the simulation, and the size effect can just be ignored. Before the pressure is applied to the ZnO nanocrystals, the nanocrystals are relaxed for 20 ps at zero pressure. The pressure is increased from 0 GPa with pressure increasing in steps of 2 GPa until the structural transition is completed, allowing 20 ps equilibration for each pressure increment, then the trajectories are collected for analyses.

Using reliable interatomic potential to describe the interaction between atoms of the nanocrystals is very important to the success of MD simulation. In present MD simulation, a Buckingham-type interatomic potential of form

$$E(r_{ij}) = \frac{q_i q_j}{r_{ij}} + A\exp\left(\frac{-r_{ij}}{\rho}\right) - \frac{C}{r_{ij}^6} \quad (1)$$

is applied to describe the interaction between atoms in the nanocrystal. *E* is the pair potential energy contributed by the interaction between the *i*th and *j*th ions with a distance of $r_{ij}$, $q_i$ is the charge of the *i*th ion, and the parameters of *A*, $\rho$ and *C* are constants listed in Table 1. The first term in Eq. 1 describes the long-range Coulomb interactions between two ions, the second and the third terms represent their short-range interactions. This potential effectively predicts surface properties such as surface energies,[46] which is very important in the simulations for nanocrystals due to the significant high surface to volume effect.

## 3. Results and discussion

The pressure-driven structural transition of the faceted $Zn_{560}O_{560}$ nanocrystal owning intact structure is studied concretely. The structural transition of the ZnO nanocrystal under hydrostatic pressure can be revealed by the volume-pressure relation. As shown in Fig. 2, the $Zn_{560}O_{560}$ nanocrystal volume decreases smoothly with increasing pressure up to 6 GPa, where there is an obvious drop in the volume. As the pressure increases to about 12 GPa, the volume decreases smoothly with increasing pressure again. The volume decrease of the first structural transition is about 1.2 Å$^3$. At about 26 GPa, the volume decreases abruptly. This means that the $Zn_{560}O_{560}$ nanocrystal would experience structural transition twice before the whole transition process is completed.

In order to quantitatively describe the transition, we have calculated the variations of coordination with pressure, which is shown in Fig. 3. Here, the atoms are defined to be the nearest neighbors if the distance between them is less than 2.6 Å. We could see that at atmospheric pressure, besides fourfold coordination atoms those account for nearly 70% of the nanocrystals, there are some threefold coordination atoms for faceted nanocrystals as result of the initial defects at the nanocrystal surface. As the pressure increases, the number of fourfold coordination and threefold coordination atoms remains almost unchanged for a certain range of pressure. But when the pressure increases to be about 7.0 GPa, the number of threefold and fourfold coordination atoms decreases sharply, while the number of the fivefold coordination atoms increases very fast. It means that the nanocrystal is experiencing some kind of structural transition there. The fivefold coordination atoms of the intermediate structure is dominant, so the intermediate structure may be a kind of fivefold coordination structure.

Further we also analyze intermediate structure by the snapshots of the MD simulation. As shown in Fig. 4, we find the nanocrystal transforms to a kind of fivefold coordination structure by a simple axial compression in the *c* direction, which is stabilized by a favorable surface free energy.[47] From the visual inspection of the MD simulation we can see the B4 to fivefold coordination structural transition of the $Zn_{560}O_{560}$ always starts on one side of the crystal compressed along the wurtzite *c* axis. This kind of fivefold coordination structure has been reported as a stable phase of MgO under hydrostatic tensile loading.[48, 49] However, for the ZnO faceted nanocrystal, this kind of structure is just metastable. When the pressure continues to increase, the fivefold coordination structure will transform to sixfold coordination B1 structure at about 26.0 GPa, which can be easily identified through the change from predominant fivefold coordination to sixfold coordination. The fivefold coordination structure to B1 structural transition is formed through changing of bond angle from 120 degree to 90 degree in the (001) plane with nucleation on the nanocrystal surface and growth by sliding of parallel crystal planes.

We also simulate several similar size faceted ZnO nanocrystals with atoms removed on the surface to study the effect on the transition path by the surface morphology. Fig. 5 shows the structure of faceted ZnO nanocrystal with atoms removed after being relaxed at 300 K and 0 GPa. It can be seen that the morphology of the local region around the surface with atoms removed becomes disordered. As the number of atoms removed at the surface edges increases, the nanocrystals become more stable and the local morphology is more similar to spherical. We therefore continue to analyze the effect on structural transition path by the disordered surface morphology. As shown in the Fig. 6, the $Zn_{555}O_{555}$ experiences a structural transition at about 5 GPa and another one at about 28 GPa, this is similar to the $Zn_{560}O_{560}$. The volume decrease of first structural transition is about 0.7 $Å^3$, smaller than that of the $Zn_{560}O_{560}$. For $Zn_{551}O_{551}$ nanocrystal, the twice structural transition can also be observed in the simulation as shown in Fig. 6. But the volume decrease of first structural transition is just 0.4 $Å^3$. For $Zn_{547}O_{547}$ with most atoms removed on the surfaces does not experience the structural transitions twice. It can be seen from Fig. 6 that the $Zn_{547}O_{547}$ nanocrystal volume decreases smoothly with increasing pressure up to the critical pressure of 25 GPa, only one structural transition can be observed.

The variations of coordinations with pressure are also calculated to identify the differences of above nanocrystals more clearly. As shown in Fig. 7, the trend of relative coordination for $Zn_{555}O_{555}$ and $Zn_{551}O_{551}$ nanocrystal are similar to the trend of the $Zn_{560}O_{560}$ nanocrystal. With a more careful comparison, we find that the relative coordination of fivefold atoms for the $Zn_{555}O_{555}$ and $Zn_{551}O_{551}$ nanocrystal is less than that of the $Zn_{560}O_{560}$ and the relative coordination of fourfold atoms is more. The relative coordination of fivefold atoms for $Zn_{555}O_{555}$ is about 0.5 and that of $Zn_{551}O_{551}$ nanocrystal is about 0.35, while that of $Zn_{560}O_{560}$ is about 0.6. The fourfold coordination B4 structure transforms to the B1 structure directly for $Zn_{547}O_{547}$ with no fivefold coordination structure as intermediate status. However the fivefold relative coordination of $Zn_{547}O_{547}$ B1 atoms accounting for about 0.3 is just because of the initial defects on the nanocrystal surface. The intermediate structures at 18 GPa of the $Zn_{555}O_{555}$ and the $Zn_{551}O_{551}$ nanocrystal are shown in Fig. 8(a) and (b), we find that not the whole nanocrystal transform to fivefold coordination structure and the fivefold structure ratio of the $Zn_{555}O_{555}$ nanocrystal is larger than that of the $Zn_{551}O_{551}$ nanocrystal. Less ratio of fivefold coordination structure reflects in less volume decrease of the first structural transition. The side with disordered surface morphology of nanocrystal keeps the B4 structure. Fig. 8(c) shows the intermediate structure at 18 GPa of $Zn_{547}O_{547}$ nanocrystal, we can see the whole $Zn_{547}O_{547}$ keeps the wurtzite structure unchanged, there is no fivefold coordination structure found during the whole transition process. For $Zn_{560}O_{560}$, $Zn_{555}O_{555}$, $Zn_{551}O_{551}$ and $Zn_{447}O_{447}$ nanocrystal, as the area of disordered surface morphology become more, the ratio of B4 structure transforming to the fivefold coordination structure during the first structural transition becomes less. The B4 to fivefold coordination structure of the ZnO nanocrystal always starts on one side with perfect structure, so it will be more difficult to transform to the fivefold coordination structure for B4 structure if there are more disordered surface morphology. The reason may lie in that the nanocrystal with disordered surface morphology will receive equal stress from every direction as spherical one and it will be difficult to compress the structure along only $c$ axis.

To verify our assumption we have also studied the structural transition of the spherical $Zn_{502}O_{502}$ nanocrystal to compare with the faceted ones. The variation of coordinations with pressure of spherical $Zn_{502}O_{502}$ nanocrystal is shown in Fig. 9. We can see the spherical $Zn_{502}O_{502}$ nanocrystal transforms from B4 to B1 structure directly, which is indicated by the direct transition from fourfold to sixfold coordination structure. Due to the symmetry, the spherical nanocrystal cannot be compressed in any special direction, such as the $c$ axis.

It is different from actual experiments on ZnO nanocrystals that the fivefold coordination structure is not observed in experiments. Firstly, a possibly metastable fivefold coordination structure might have a lifetime too short to be resolved in the experiments. Secondly, crystals used in experiments are covered with surfactants, which strongly influence surface energies.[50] The presence of such a surface passivation layer may block the transition path to the fivefold coordination structure. Our simulation reveals that maybe the disordered surface morphology of surface can also change the transition path and makes no intermediate structure occurring. While there is usually no perfect morphology surfaces in the ZnO nanocrystals synthesized in experiment, this may be another reason that the fivefold coordination structure could not be observed in the experiments.

## 4. Conclusions

In summary, the faceted ZnO nanocrystal with perfect surface structure experiences twice structural transition during the B4 to B1 structural transition and the whole nanocrystal transforms to the perfect fivefold coordination structure during the first structural transition. But for the faceted ones with disordered surface, the local morphology is more similar to spherical as the area of disordered surface increases. When the disordered surface become more, the nanocrystal will receive equal stress from every direction as spherical nanocrystal and it will be more difficult to compress the structure along only $c$ axis. As a result only partial structure experiences intermediate fivefold coordination structure or even no intermediate fivefold coordination structure exists dependent on the surface disorder level.

## Acknowledgments

This work is supported by the National Natural Science Foundation of China [No. 11004135] and Innovation Program of Shanghai Municipal Education Commission [No. 11YZ84].

# Figure captions

**FIG. 1** The structure of the $Zn_{560}O_{560}$ nanocrystal is shown in the left, the $(001)$ and $(00\bar{1})$ surfaces of the $Zn_{560}O_{560}$ nanocrystal are shown in the right, O atoms are colored by red and Zn atoms by cinereous. The $Zn_{555}O_{555}$ nanocrystal is obtained by removing atoms labeled 1 of $Zn_{560}O_{560}$ nanocrystal, the $Zn_{551}O_{551}$ nanocrystal is obtained by removing atoms labeled 1 and 2 of $Zn_{560}O_{560}$ nanocrystal, the $Zn_{547}O_{547}$ nanocrystal is obtained by removing atoms labeled 1, 2 and 3 of $Zn_{560}O_{560}$ nanocrystal.

**FIG. 2** Volume versus pressure for faceted $Zn_{560}O_{560}$. The discontinuity of the slopes indicates structural transition under hydrostatic pressure. It can be seen that $Zn_{560}O_{560}$ experiences structural transition twice through the whole transition process.

**FIG. 3** Relative coordination $N_i/N$ versus pressure for faceted $Zn_{560}O_{560}$. The coordination number is defined as the number of the nearest neighbor atoms.

**FIG. 4** The snapshots of the MD simulation for the faceted $Zn_{560}O_{560}$ nanocrysal transforming from B4 structure to B1 structure. (a), (b) and (c) show the side view of wurtzite structure, side view of fivefold coordination structure at pressure of 18 GPa and the cross-sectional of final rocksalt structure respectively.

**FIG. 5** The local regions morphology around surface edges with atoms removed become disordered shown by the yellow atoms. The perfect surface structure is also shown for comparison.

**FIG. 6** Volume versus pressure for faceted $Zn_{555}O_{555}$, $Zn_{551}O_{551}$ and $Zn_{547}O_{547}$.

**FIG. 7** Relative coordination $N_i/N$ versus pressure for faceted $Zn_{555}O_{555}$, $Zn_{551}O_{551}$ and $Zn_{547}O_{547}$. The coordination number is defined as the number of the nearest neighbor atoms.

**FIG. 8** The snapshots of the MD simulation for the faceted nanocrysals transition from B4 to B1 structure. (a), (b) and (c) show the side view of the structure at pressure of 18 GPa for $Zn_{555}O_{555}$, $Zn_{551}O_{551}$ and $Zn_{547}O_{547}$.

**FIG. 9** Relative coordination $N_i/N$ versus pressure for spherical $Zn_{502}O_{502}$.

**Table 1** Short-range interaction parameters for ZnO.

| Species | $A$(eV) | $\rho$(Å) | $C$(eV Å) |
|---|---|---|---|
| $O^{2-}O^{2-}$ | 9547.96 | 0.21916 | 32.0 |
| $Zn^{2+}O^{2-}$ | 529.70 | 0.3581 | 0.0 |
| $Zn^{2+}Zn^{2+}$ | 0.0 | 0.0 | 0.0 |

# Figures

**FIG. 1**

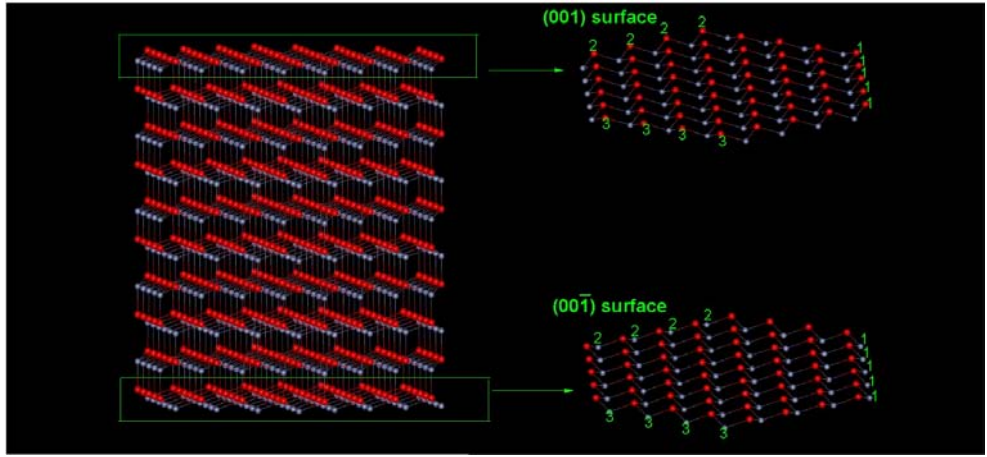

**FIG. 2**

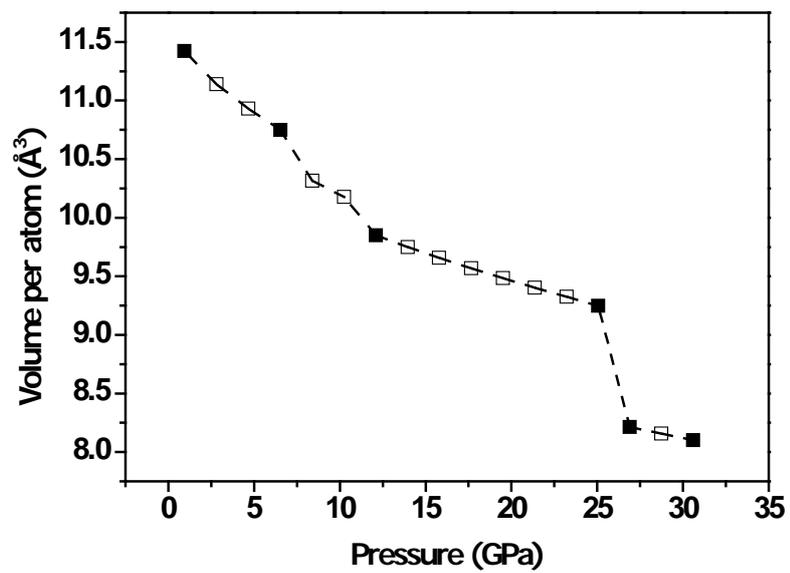

**FIG. 3**

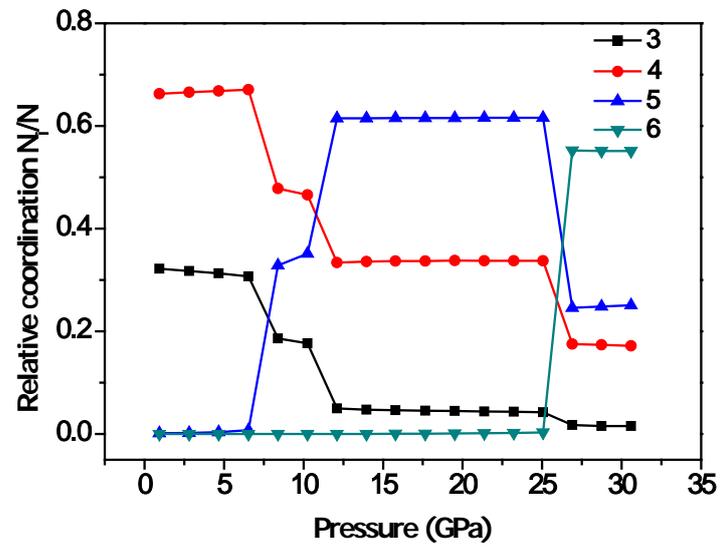

**FIG. 4**

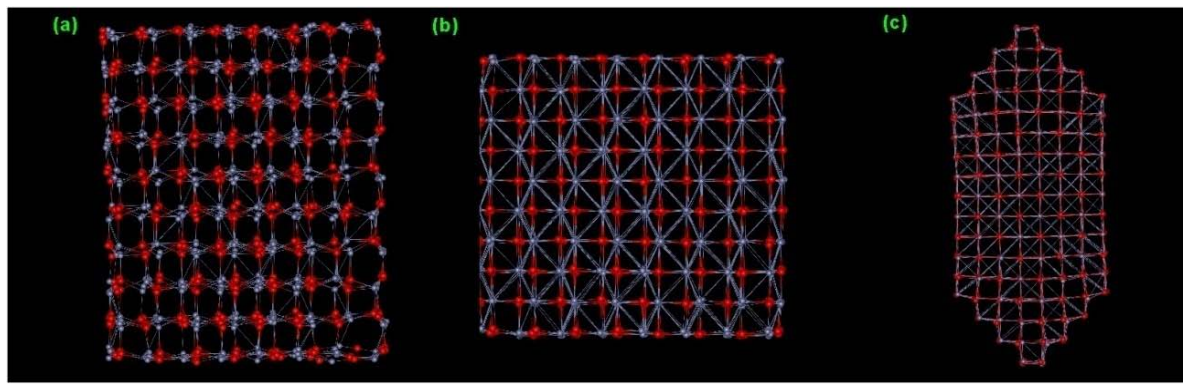

**FIG. 5**

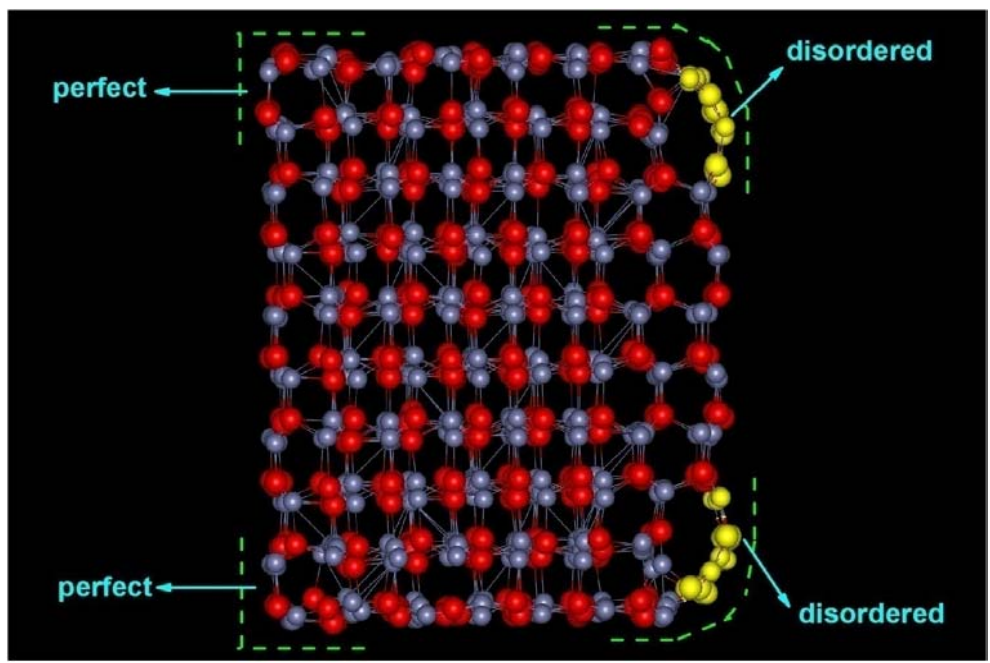

**FIG. 6**

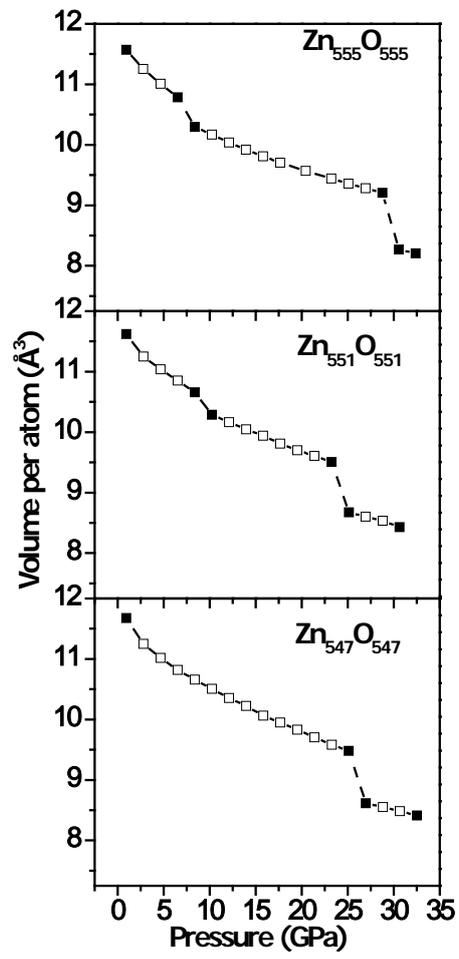

**FIG. 7**

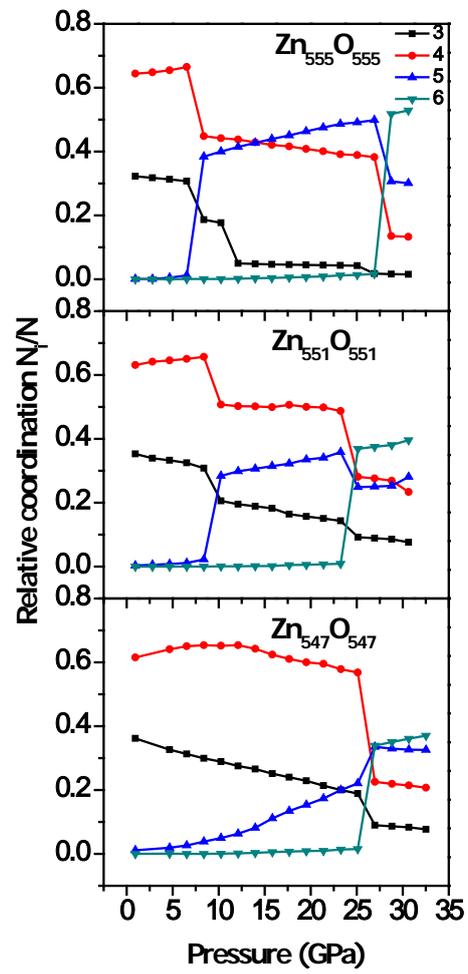

**FIG. 8**

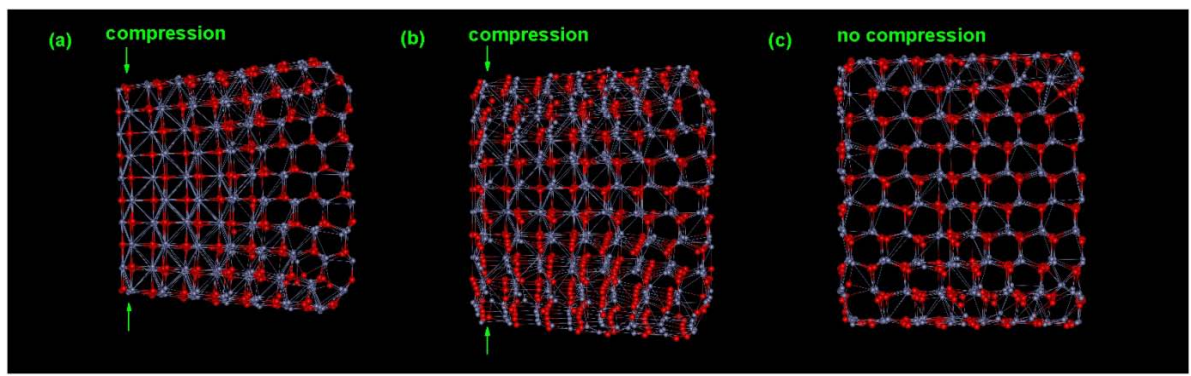

**FIG. 9**

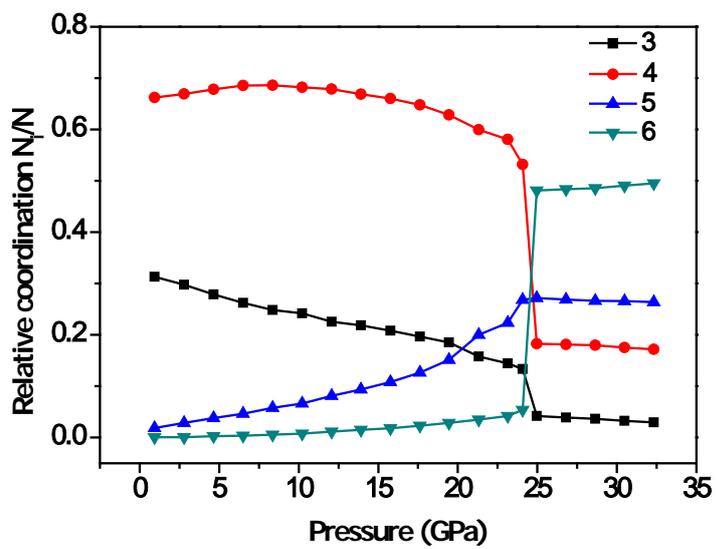